\documentclass[epj]{webofc}

\usepackage[normalem]{ulem}
\usepackage{graphicx,color}
\usepackage[varg]{txfonts}   

\wocname{EPJ Web of Conferences}
\woctitle{INPC 2013}

\newcommand{\avep}{AV8$^{\prime}$~}

\begin{document}

\title{From nuclear structure to neutron stars}
\author{Stefano Gandolfi\inst{1}
\and
Andrew W. Steiner\inst{2}
}
\institute{Theoretical Division, Los Alamos National Laboratory
Los Alamos, NM 87545, USA
\and
Institute for Nuclear Theory, University of Washington, Seattle, WA 98195
}

\abstract{ 
    Recent progress in quantum Monte Carlo with modern nucleon-nucleon
    interactions have enabled the successful description of properties
    of light nuclei and neutron-rich matter. As a demonstration, we
    show that the agreement between theoretical calculations of the
    charge form factor of $^{12}\mathrm{C}$ and the experimental data is
    excellent. Applying similar methods to isospin-asymmetric systems
    allows one to describe neutrons confined in an external potential
    and homogeneous neutron-rich matter. Of particular interest is the
    nuclear symmetry energy, the energy cost of creating an isospin
    asymmetry. Combining these advances with recent observations of
    neutron star masses and radii gives insight into the equation of
    state of neutron-rich matter near and above the saturation
    density. In particular, neutron star radius measurements constrain
    the derivative of the symmetry energy.}
\maketitle

\section{Introduction}

In the last few decades, properties of nuclear systems have been
successfully described by nucleon-nucleon potentials like Argonne and
Urbana/Illinois forces, that reproduces two-body scattering and
properties of light nuclei with very high
precision~\cite{Wiringa:1995,Pieper:2001}. These nuclear potentials
reproduce several properties of nuclear systems extremely well, including
binding energies of ground- and excited states, radii, matrix
elements, scattering states, and other
observables~\cite{Pieper:2008,Pieper:2001b,Nollett:2007,Schiavilla:2007}.

The Argonne AV18 nucleon-nucleon interaction has small non-local
terms and a hard core. 
Direct diagonalization of the
Hamiltonian is not possible, and the expansion of the wave function on
a finite basis, for example using no-core shell model or couple
cluster methods, is computationally very expensive or unfeasible.
In contrast, the use of correlated wave functions combined with
Quantum Monte Carlo (QMC) methods has  provided highly accurate solutions of the ground state 
of many-body nuclear systems~\cite{Pudliner:1997}.

The knowledge of the Equation of State (EoS)
of pure neutron matter is an important bridge from the
  nucleon-nucleon interaction to neutron-rich matter. 
The symmetry
energy $E_{\rm sym}$ is the difference of nuclear matter and neutron
matter energy and gives the energy cost of the
isospin-asymmetry in the homogeneous nucleonic matter. In the last few
years the study of $E_{\rm sym}$ has received considerable attention
(see for example Ref.~\cite{Tsang:2012} for a recent
experimental/theoretical review). The role of the symmetry energy is
essential to understand the mechanism of stability of very
neutron-rich nuclei, and is also related to many
phenomena occurring in neutron stars. The number of protons per
  baryon, $x$, is determined by beta-equilibrium and charge
  neutrality. These imply relationships between the chemical
  potentials and the symmetry energy, $\mu_e = \mu_n - \mu_p \approx 4
  E_{\mathrm{sym}} (1-2 x) $. Matter near the nuclear saturation
  density is very neutron-rich, because electron degeneracy drives
  $\mu_n > \mu_p$. Thus neutron star matter is
  sensitive to $E_{\rm sym}$ and
its first derivative. 
  The inner
crust of neutron stars, where the density is a fraction of nuclear
densities, is mostly composed of neutrons surrounding a matter made of
extremely-neutron rich nuclei that, depending on the density, may
exhibit very different phases and properties. The extremely rich phase
diagram of crustal matter is strongly related to
the role of $E_{\rm sym}$. For example, it governs the
phase-transition between the crust and the core~\cite{Newton:2011} and
the nature of $r$-mode 
instabilities~\cite{Wen:2012,Vidana:2012}.

  Neutron drops, neutrons confined by an external
  potential,
  provide a very simple model of neutron-rich nuclei, in
which the core is modeled as an external potential acting on valence
neutrons. In
Refs.~\cite{Chang:2004,Pieper:2005,Gandolfi:2006} neutron-rich oxygen
isotopes were successfully described by neutrons
confined in external wells, and in Ref.~\cite{Gandolfi:2008}, the same
model has been used to study calcium isotopes. More importantly,
these systems describe inhomogeneous neutron matter that can be used
as data for calibrating model energy density functionals in several
conditions~\cite{Gandolfi:2011,Bogner:2011}. The use of these
functionals to study nuclei close to the neutron drip line requires
then an important extrapolation to large isospin-asymmetries. This
extrapolation is even more dramatic when the Skyrme forces are used to
study the properties of the neutron star crust, where the matter is
made by extremely neutron-rich nuclei surrounded by a sea of neutrons.
For these reasons, \emph{ab-initio} calculations of these systems
starting from accurate nuclear Hamiltonians are important to constrain
density functionals.

\section{The Nuclear Hamiltonian and Quantum Monte Carlo }

In our model, neutrons are non-relativistic point-like particles
interacting via two- and three-body forces:
\begin{equation}
H = \sum_{i=1}^A\frac{p_i^2}{2m} + \sum_{i<j}v_{ij}+\sum_{i<j<k}v_{ijk} \,.
\end{equation}
The two body-potential that we use is the Argonne
\avep\cite{Wiringa:2002}, that is a simplified form of the Argonne
AV18~\cite{Wiringa:1995}. Although simpler to use in QMC calculations,
\avep provides almost the same accuracy as AV18 in fitting NN
scattering data~\cite{Gandolfi:2013}. The three-body force is not as
well constrained as the NN interaction, but its inclusion in realistic
nuclear Hamiltonians is important to correctly describe the binding
energy of light nuclei~\cite{Pieper:2001}.

The Urbana IX (UIX) three-body force has been originally proposed in
combination with the Argonne AV18 and \avep\cite{Pudliner:1995}.
Although it slightly underbinds the energy of light nuclei, it has
been extensively used to study the equation of state of nuclear and
neutron matter~\cite{Akmal:1998,Gandolfi:2009,Gandolfi:2012}. The
Illinois forces have been introduced to improve the description of
both ground- and excited-states of light nuclei, showing an excellent
accuracy~\cite{Pieper:2001,Pieper:2008}, but it produces an unphysical
overbinding in pure neutron systems~\cite{Sarsa:2003,Maris:2013}.

Another interesting class of nucleon-nucleon potentials are derived
within the chiral effective field theory. Typically, these
interactions have strong non-local terms, and as a consequence they
cannot be easily included in QMC calculations. Recently it has been
showed that these potentials can be designed to be local, and combined
with QMC simulations~\cite{Gezerlis:2013}. However, the need to
include a cutoff to the nucleon's momentum limits
the applicability of chiral forces to study dense neutron matter. The
cutoff of these potentials can be controlled in a many-body
calculation~\cite{Gezerlis:2013}, but the uncertainty is already quite
large at saturation density in neutron matter, making the calculation
at larger densities unfeasible.

We solve the many-body ground-state with a projection in imaginary-time,
i.e.:
\begin{equation}
\Psi(\tau)=\exp[-H\tau]\Psi_v \,,
\end{equation}
where $\Psi_v$ is a variational ansatz, and $H$ is the Hamiltonian of
the system. In the limit of $\tau\rightarrow\infty$, $\Psi$ approaches
the ground-state of $H$. The evolution in imaginary-time is performed
by sampling configurations of the system using Monte Carlo techniques,
and expectation values are evaluated over the sampled configurations.
The main difference between GFMC and AFDMC is in the way that
spin/isospin states are treated. In GFMC, all the spin/isospin states
are explicitly included in the variational wave function. The results
obtained are very accurate but limited to the
$^{12}\mathrm{C}$~\cite{Pieper:2008} or 16 neutrons~\cite{Gandolfi:2011}. The
AFDMC method samples the spin/isospin states using the
  Hubbard-Stratonovich transformation rather than simpling them
  explicitly~\cite{Schmidt:1999}. The calculation can be then
extended up to many neutrons, making the simulation of homogeneous
matter and heavy nuclear systems possible~\cite{Lonardoni:2013}. 
The AFDMC has proven to be very accurate when
compared to GFMC calculation of energies of neutrons confined in an
external potential~\cite{Gandolfi:2011}. We shall present results
obtained either using GFMC and AFDMC.

\section{The Form Factor of 
$^{12}\mathrm{C}$}

The spectra of light nuclei has been calculated using GFMC. Several
ground- and excited-states are reproduced with high accuracy, with an
average deviation with respect experimental measurements of the order
of few keV for nuclei from deuteron up to $^{12}\mathrm{C}$. We have recently
calculated also the form factor of $^{12}\mathrm{C}$ and
  find excellent agreement with a compilation of experimental
data~\cite{Lovato:2013}.

The charge (and currents) operators are generally written by including 
one- and two-body operators:
\begin{equation}
\rho_q=\sum_i\rho_q(i)+\sum_{i<j}\rho_q(ij) \,.
\end{equation}
The above operators are described for example in Ref.~\cite{Shen:2012}

\begin{figure}
\begin{center}
\includegraphics[width=0.7\textwidth]{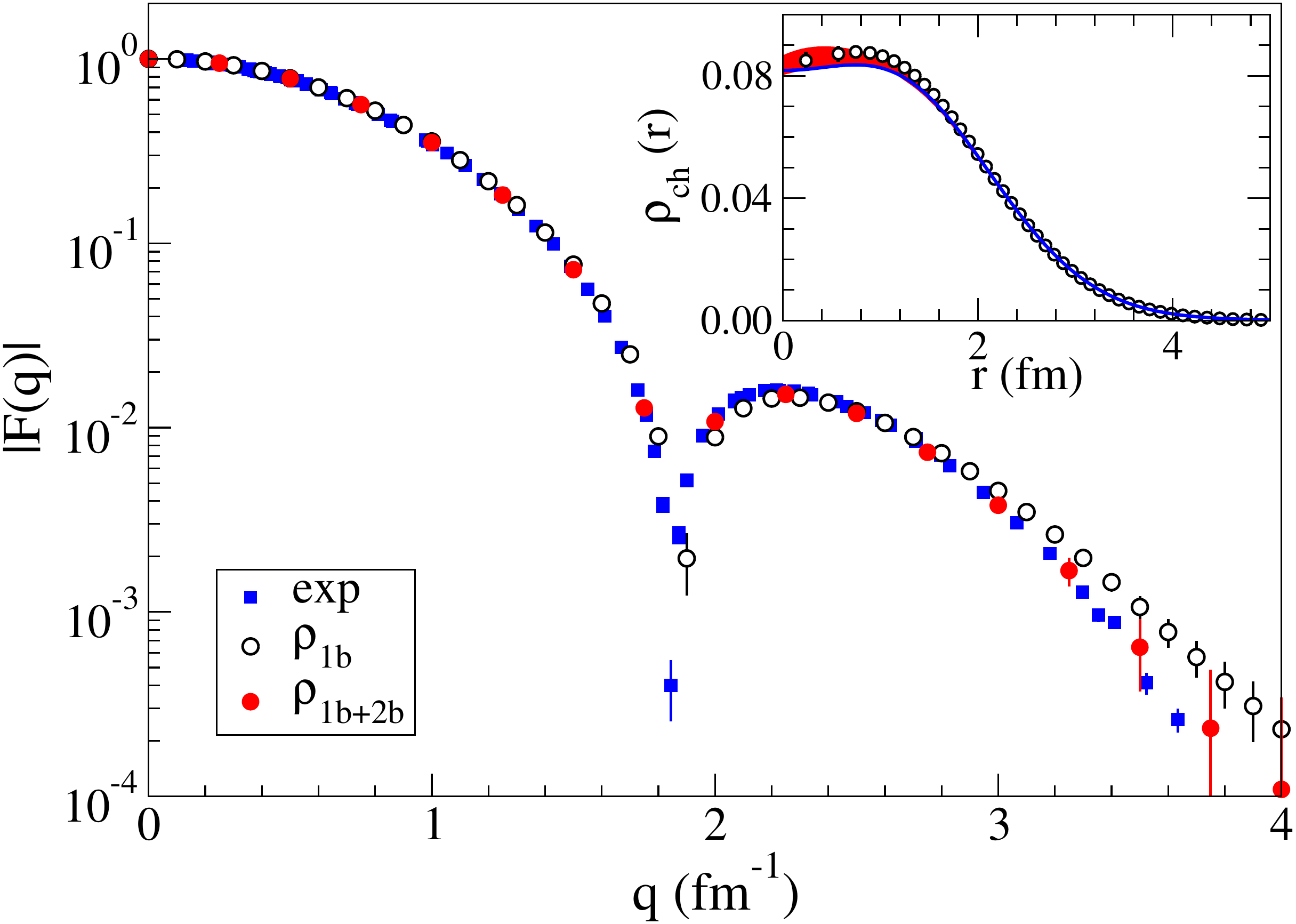}
\caption{The form factor of $^{12}\mathrm{C}$ calculated using GFMC with the
  AV18+IL7 Hamiltonian. The blue squares are experimental data, open
  black circles are the results obtained using only one-body charge
  operator and full red circles are the obtained with one- and
  two-body operators. In the inset we show the charge distribution in
  the nucleus, obtained with a Fourier transform of $F(q)$. The figure
  is taken from Ref.~\cite{Lovato:2013}. }
\label{fig:ffc12}
\end{center}
\end{figure}

In Fig.~\ref{fig:ffc12} we show the GFMC results of the form factor.
In this case the role of two-body operators is appreciable only for
high momentum transfer $q\ge 3$ fm$^{-1}$. In all the range of momenta
considered, the agreement between the calculation and experimental
data is excellent. The same operators have been employed to calculate
the electro-magnetic sum-rules of $^{12}\mathrm{C}$. Especially in the
transverse sum-rule, the two-body operators (that are commonly
neglected in similar calculations) contributes up to 50\%. The large
contribution given by the two-body currents has also been showed in
early calculations of the Euclidean response in
$^4$He~\cite{Carlson:2002}.

\section{Neutron Drops}

\begin{figure}
\begin{center}
\includegraphics[width=0.7\textwidth]{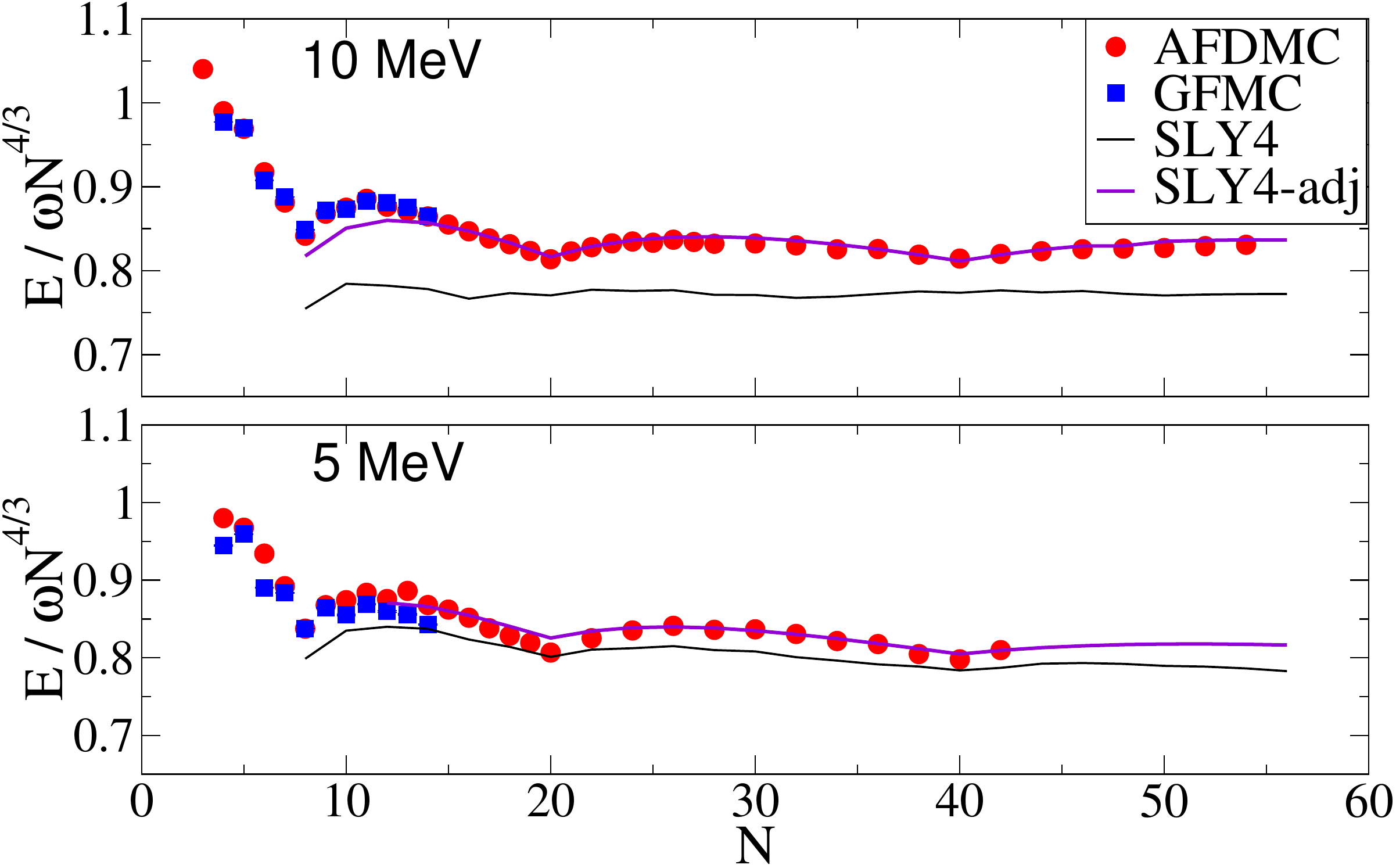}
\caption{The energy of neutrons in a HO well with $\hbar\omega=10$ MeV
  (upper panel) and 5 MeV (lower panel) in units of $\omega N^{4/3}$.
  The red dots are the results given by AFDMC, blue squares are from
  GFMC, and the black line is the result obtained using the Skyrme
  SLy4. The violet line is the adjusted SLy4 where the strength of the
  gradient, pairing, and spin-orbit terms have been changed. The
  figure is taken from Ref.~\cite{Gandolfi:2011}. }
\label{fig:eho}
\end{center}
\end{figure}

In the last few years, the energy and other properties of neutron
drops have been studied by using ab-initio
methods~\cite{Gandolfi:2011,Maris:2013} by confining neutrons in a
harmonic oscillator (HO) or in a Wood-Saxon (WS) well. The QMC energy
of neutron drops confined by $V_{HO}$ is shown in Fig.~\ref{fig:eho}
for two different frequencies of the external potential. The red
points are the results obtained using the AFDMC method, and the blue
ones using the GFMC. The two solid lines are the results given by
using the original Skyrme SLy4 force~\cite{Chabanat:1995}, and a
modified version. The energy is in units of the Thomas-Fermi energy,
that is proportional to $\omega N^{4/3}$, to see the extrapolation to
the thermodynamic limit. The two QMC methods agree within 1\% for the
$\hbar\omega=10$ MeV trap, and the difference increases up to 4\% for
$\hbar\omega=5$ MeV. The larger difference between GFMC and AFDMC
  for larger values of $\hbar\omega$ comes from the lack of pairing
  correlations in the AFDMC.
  At low densities neutrons are superfluids, and pairing
correlations are quite important to include for open-shell
configurations.

\begin{figure}
\begin{center}
\includegraphics[width=0.7\textwidth]{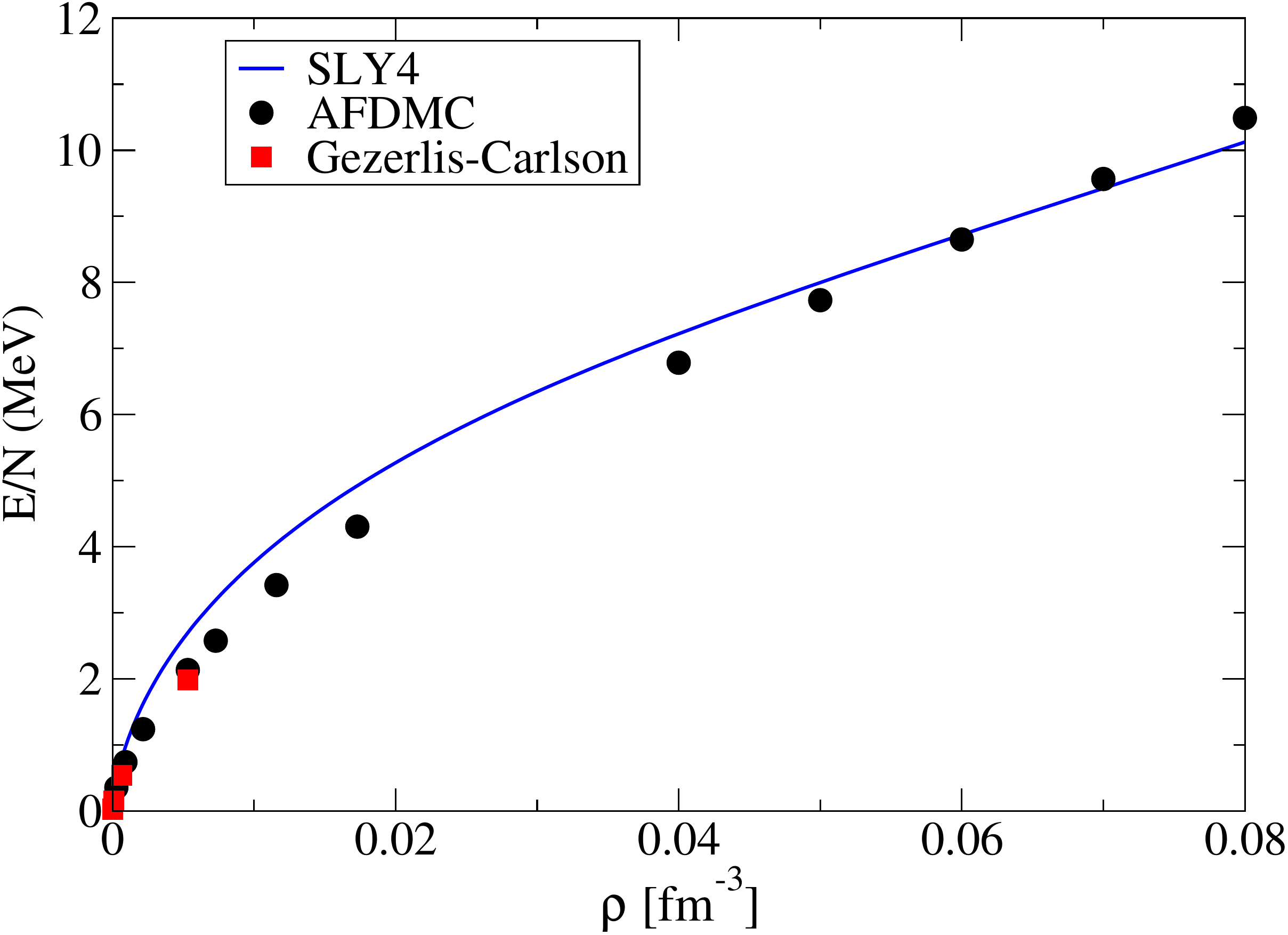}
\caption{The equation of state of pure neutron matter. The AFDMC
  results~\cite{Gandolfi:2009,Gandolfi:2008b} are compared to the QMC
  results at low densities of Gezerlis and
  Carlson~\cite{Gezerlis:2008,Gezerlis:2010}, and to the equation of
  state of Skyrme SLy4~\cite{Chabanat:1998}. }
\label{fig:eossk}
\end{center}
\end{figure}

The difference between QMC and Skyrme at closed shells is mainly due
to two effects, the bulk contribution and the gradient term. Skyrme
forces typically give an EoS of pure neutron matter at densities lower
than saturation that is more repulsive than microscopic calculations.
The equation of state of pure neutron matter is shown in
Fig.~\ref{fig:eossk} where we compare the AFDMC results from
Refs.~\cite{Gandolfi:2009,Gandolfi:2008b,Gandolfi:2009b}, the GFMC
calculation of Gezerlis and
Carlson~\cite{Gezerlis:2008,Gezerlis:2010}, and the equation of state
given by SLy4. We make the reasonable assumption that Skyrme's bulk
term cannot explain the difference between QMC and Skyrme energy in
neutron drops. Then, since the pairing and the spin-orbit terms are
expected to be very weak with respect to the gradient term for closed
shell configurations, we can use the energy at N=8, 20 and 40 to
re-adjust the gradient term of Skyrme. The energy of neutron drops
with N near closed shells can be used to adjust the spin-orbit
strength because for these configurations the pairing is not
important. Finally, by comparing the energy of half-filled shells, we
can tune the pairing term. In addition to the energy of neutrons in a
HO potential, the adjusted SLy4 reproduces the energies in a WS well,
radii and radial densities~\cite{Gandolfi:2011}.

\section{The Equation of State of Neutron Matter}

In this section we present QMC results for pure neutron matter. There
are several reasons to focus on pure neutron matter. First, the
three-body interaction is non-zero only in the $T=3/2$ isospin-channel
($T$ is the total isospin of three-nucleons), while in the presence of
protons there are also contributions in $T=1/2$. The latter term is
the dominant one in nuclei, and only weakly accessible by studying
properties of nuclei. Second, the EoS of pure neutron matter is
closely related to the structure of neutron stars.

\begin{figure}
\begin{center}
\includegraphics[width=0.7\textwidth]{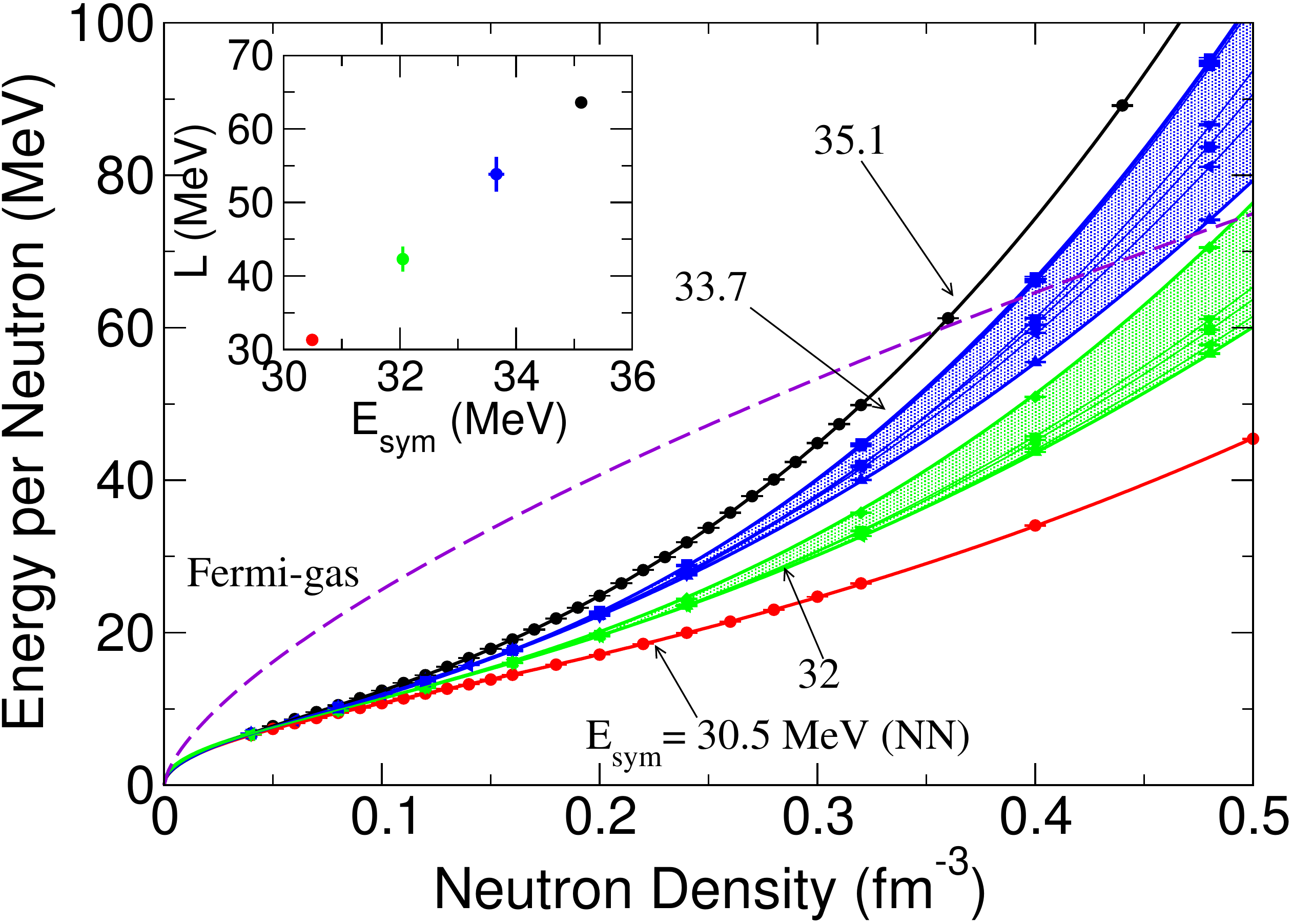}
\end{center}
\caption{The QMC equation of state of neutron matter for various
  Hamiltonians. The red (lower) curve is obtained by including the NN
  (Argonne \avep) alone in the calculation, and the black one is
  obtained by adding the Urbana IX three-body force. The green and
  blue bands correspond to EoSs giving the same $E_{\rm sym}$ (32 and
  33.7 MeV respectively), and are obtained by using several models of
  three-neutron force. In the inset we show the value of $L$ as a
  function of $E_{\rm sym}$ obtained by fitting the EoS. The figure is
  taken from Ref.~\cite{Gandolfi:2012}. }
\label{fig:eos}
\end{figure}

We present several EoSs obtained using different models of
three-neutron force in Fig.~\ref{fig:eos}. The two solid lines
correspond to the EoSs calculated using the NN potential alone and
including the UIX three-body force~\cite{Pudliner:1995}. The effect of
using different models of three-neutron force is clear in the two
bands, where the high density behavior is showed up to about
$3\rho_0$. At such high density, the various models giving the same
symmetry energy at saturation produce an uncertainty in the EoS of
about 20 MeV. The EoS obtained using QMC can be conveniently fit using
the following functional~\cite{Gandolfi:2009}:
\begin{equation}
E(\rho)=a\,\left(\frac{\rho}{\rho_0}\right)^\alpha
+b\,\left(\frac{\rho}{\rho_0}\right)^\beta \,,
\label{eq:enefunc}
\end{equation}
where $E$ is the energy per neutron, $\rho_0=0.16$ fm$^{-3}$, 
and $a$, $b$, $\alpha$ and $\beta$ are free parameters.
The parametrizations of the EoS obtained from different nuclear Hamiltonians
is given in Ref.~\cite{Gandolfi:2012}.

At $\rho_0$ symmetric nuclear matter saturates, and we can extract the
value of $E_{\rm sym}$ and $L$ directly from the pure neutron matter
EoS. The result of fitting the pure neutron matter EoS is shown in the
inset of Fig.~\ref{fig:eos}. The error bars are obtained by taking the
maximum and minimum value of $L$ for a given $E_{\rm sym}$, and the
curves obtained with NN and NN+UIX are thus without error bars. From
the plot it is clear that within the models we consider, the
correlation between $L$ and $E_{\rm sym}$ is linear and quite strong.

\begin{figure}
\begin{center}
\includegraphics[width=0.7\textwidth]{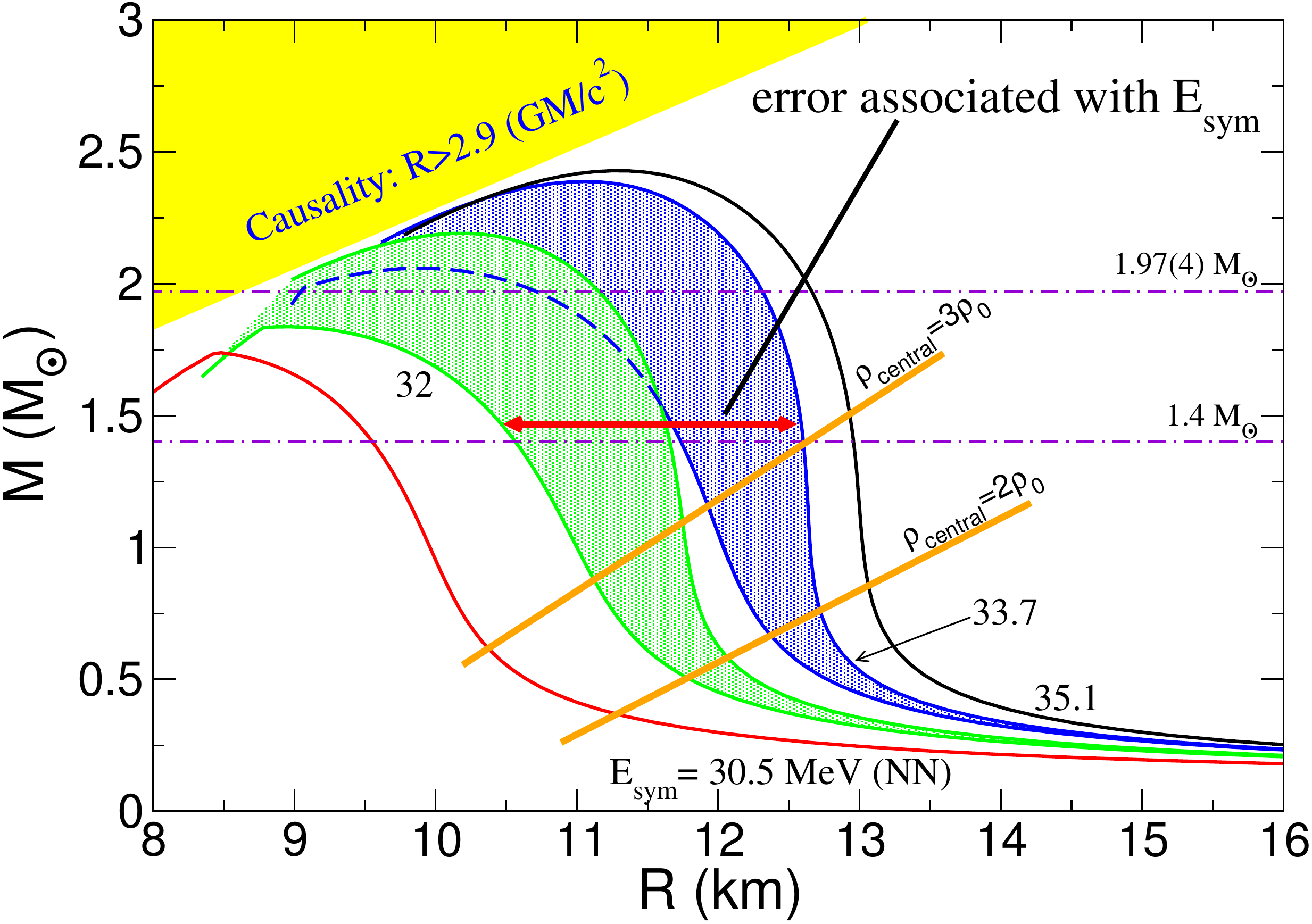}
\end{center}
\caption{The mass-radius relation of neutron stars obtained from the
  EoS calculated using QMC. The various colors represent the $M-R$
  result obtained from the corresponding EoSs described in
  Fig.~\ref{fig:eos}. The two horizontal lines show the value of
  $M=1.4$ and 1.97(4)$M_{\odot}$~\cite{Demorest:2010}. The figure is
  adapted from Ref.~\cite{Gandolfi:2012}. }
\label{fig:nstar}
\end{figure}

\section{Connection to Neutron Star Masses and Radii}

    Neutron stars, unlike planets, are expected
to be compositionally uniform, in which case their radius is
determined principally by their mass; to a good approximation all
neutron stars lie on a universal mass-radius $M-R$ curve. When the EoS
of the neutron star matter has been specified, the structure of an
idealized spherically-symmetric neutron star model can be calculated
by integrating the Tolman-Oppenheimer-Volkoff (TOV) equations.

The neutron star mass measurements which provide the strongest EoS
constraints are those which have the highest mass. Recent
observations~\cite{Demorest:2010,Antoniadis13} have found two neutron
stars with masses near 2 $M_{\odot}$. These two data points provide
some of the strongest constraints on the nature of zero-temperature
QCD above the nuclear saturation density. We begin by examining what
can be deduced about the M-R relation directly from these mass
measurements, without employing a separate model for high-density
matter. For lower densities we use the EoS of the crust obtained in
Refs. \cite{Baym:1971} and \cite{Negele:1973}. For the core, we begin
with the parameterization in Eq.~\ref{eq:enefunc}, employing maximally
stiff EoS when the QMC models violate the causality and become
superluminal. The mass of a neutron star as a function of its radius
is shown in Fig.~\ref{fig:nstar}. The two bands correspond to the
result obtained using the two sets of EoS giving the same value of
$E_{\rm sym}$ indicated in the figure. As in the case of the EoS, it
is clear that the main source of uncertainty in the radius of a
neutron star with $M=1.4M_{\odot}$ is due to the uncertainty of
$E_{\rm sym}$ rather than the model of the three-neutron force. The
addition of a small proton fraction would change the radius $R$ only
slightly~\cite{Gandolfi:2010,Akmal:1998}, smaller than other
uncertainties in the EoS that we have discussed. The numbers in the
figure indicate the symmetry energy associated with the various
equations of state. In the figure we also indicate with the orange
lines the density of the neutron matter inside the star. Even at large
masses the radius of the neutron star is mainly governed by the
equation of state of neutron matter between 1 and 2
$\rho_0$~\cite{Lattimer:2001}.

The \avep Hamiltonian alone does not support the recent observed
neutron star with a mass of 1.97(4)M$_\odot$~\cite{Demorest:2010}.
However, adding a three-body force to \avep can provide sufficient
repulsion to be consistent with all of the
constraints~\cite{Gandolfi:2012}. There is a clear correlation between
neutron star radii and the symmetry energy which determines the EoS of
neutron matter between 1 and 2 $\rho_0$. The results in
Fig.~\ref{fig:nstar} also show that the most modern neutron matter EoS
imply a maximum neutron star radius not larger than about 13 km, unless a
drastic repulsion sets in just above the saturation density. This
tends to rule out large values of $L$, typical of Walecka-type
mean-field models without higher-order meson couplings which can
decrease $L$.

\section{Radius Measurements}
\label{sec:radius}

\begin{figure}
\includegraphics[width=0.49\textwidth]{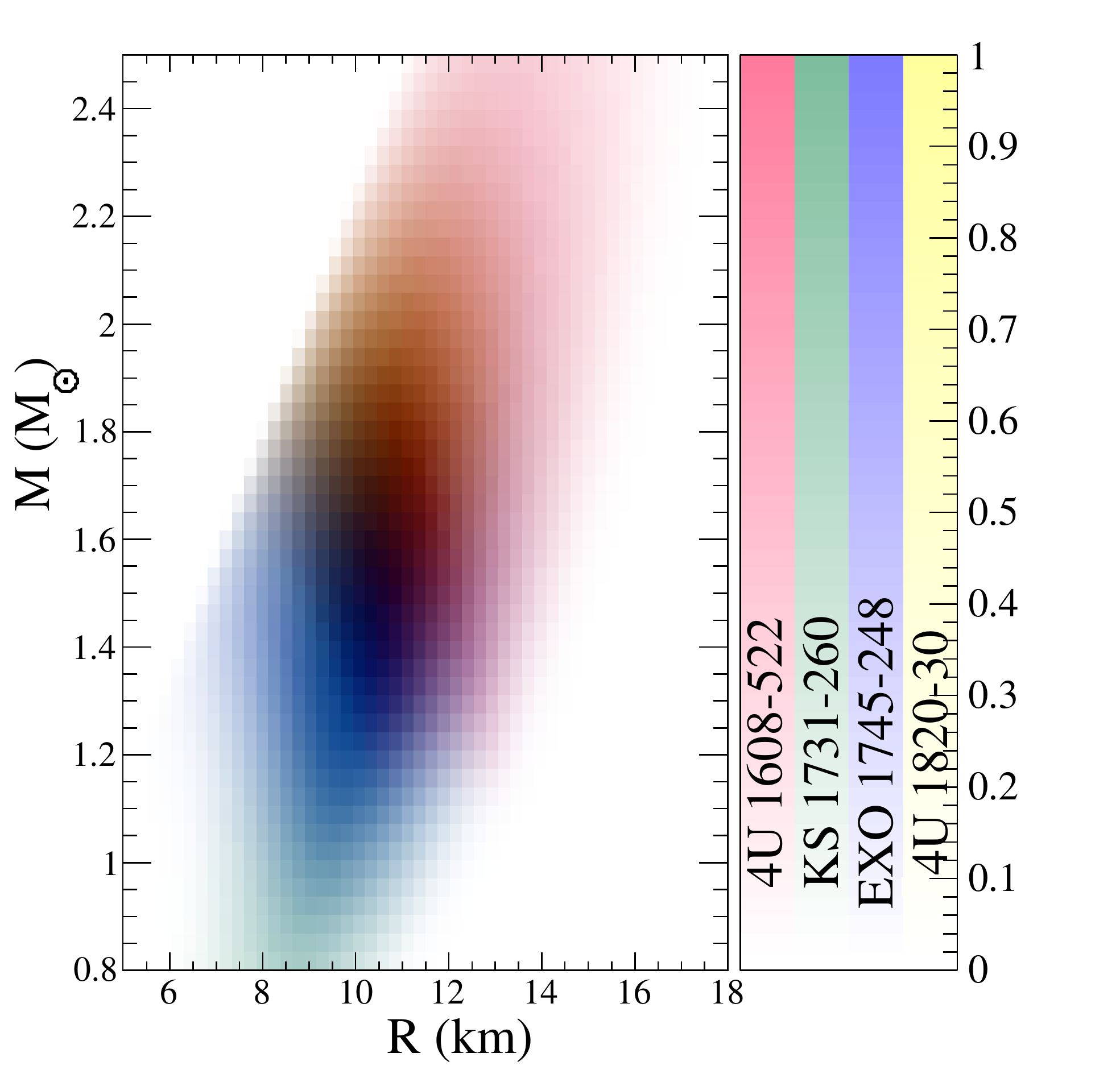}
\includegraphics[width=0.49\textwidth]{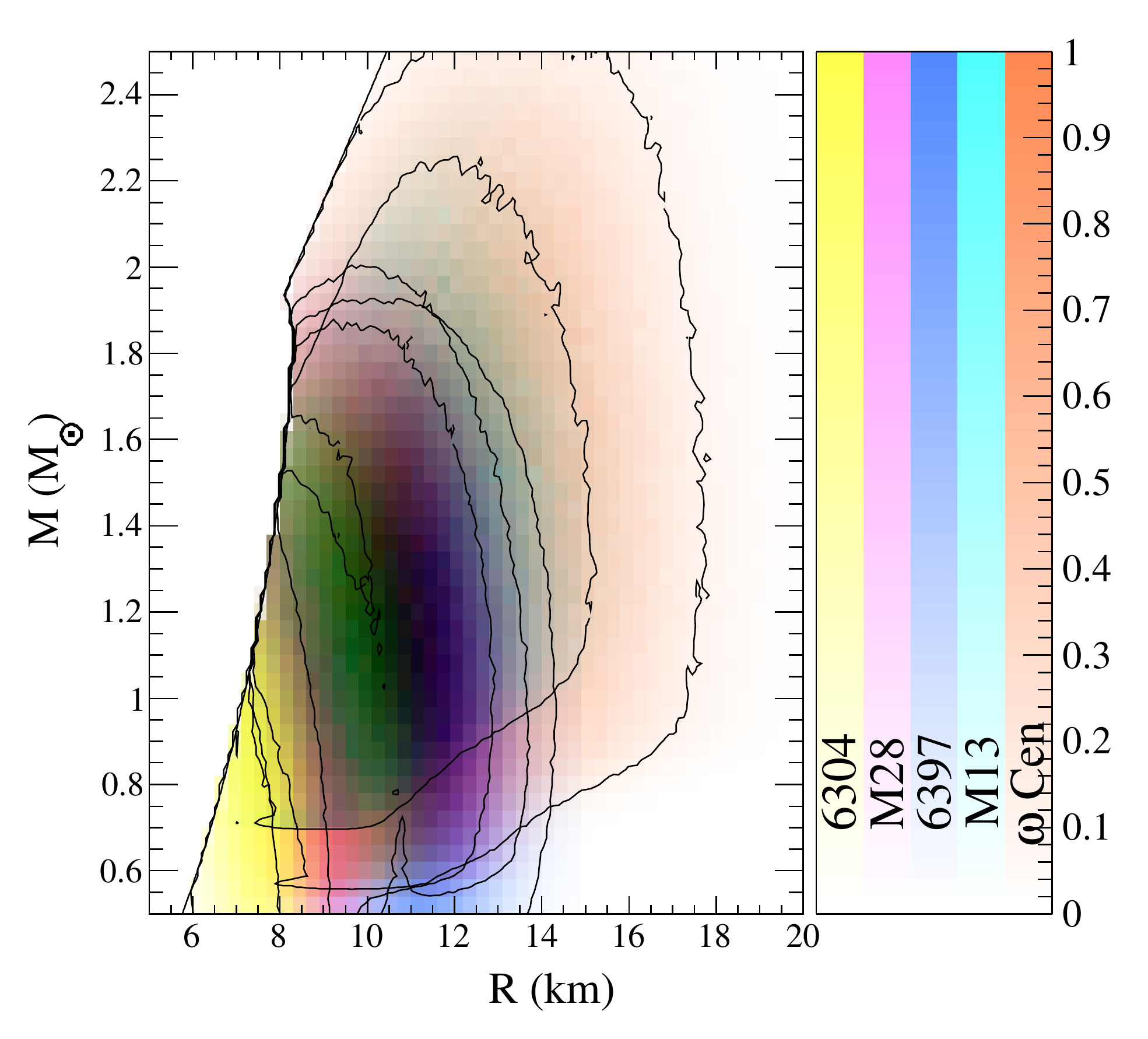}
\caption{Left panel: Probability distributions in the mass-radius
  plane for four neutron stars exhibiting PRE X-ray bursts. Colors are
  added together in RGB color space. Right panel: Probability
  distributions in the mass-radius plane for five neutron stars in
  five globular clusters from Ref.~\cite{Lattimer13}. Colors are added
  together in RGB color space when necessary. The contour lines
  outline the 90\% confidence regions.}
\label{fig:ns_rad}
\end{figure}

In contrast to the mass measurements described above, neutron star
radius measurements have proven more difficult, because they require
both a distance measurement and some degree of modeling of the neutron
star X-ray spectrum. Low-mass X-ray binaries (LMXBs) are neutron stars
accreting matter from a low mass main-sequence or white dwarf
companion. There are two types of LMXB observations which have
recently provided neutron star radius information. The first type are
LMXBs which exhibit photospheric radius expansion (PRE) X-ray bursts,
thermonuclear explosions strong enough to temporarily lift the surface
(photosphere) of the neutron star
outwards~\cite{vanParadijs:1979,Ozel:2010}. Several neutron stars have
exhibited PRE X-ray bursts and four which have have been used to infer
the neutron star mass and radius are given in the left panel of
Fig.~\ref{fig:ns_rad}, using the methods described in
Ref.~\cite{Steiner:2010}. The second type are quiescent LMXBs,
(QLMXBs), where the accretion from the companion has stopped, allowing
observation of the neutron star surface which has been heated by
accretion~\cite{Rutledge:1999}. A recent analysis of five neutron
stars~\cite{Lattimer13} including the possibility of both hydrogen and
helium atmospheres and distance uncertainties is shown in the right
panel of Fig.~\ref{fig:ns_rad}. Note that already from these two
figures alone, it is clear that these probability distributions favor
neutron star radii near 11 km. Although we will similar $(R,M)$
distributions in our analysis below, it is important to remember that
there are several systematic uncertainties which are potentially
important. For the QLMXBs, the treatment of the X-ray absorption
between the source and the observer, the flux calibration of the
observing satellite, and the method used to measure the distance all
play important roles. The situation for PRE X-ray bursts is even more
challenging: complications such as spherical asymmetry, the time
evolution of the spectra, and the location of the photosphere at
``touchdown'' may all modify the implied masses and radii.


\section{Bayesian Analysis of Neutron Star Masses and Radii}
\label{sec:bayes}

In this section, we constrain the EoS and symmetry energy using
observational data similar to that described in Sec.~\ref{sec:radius}
and include the possibility of phase transitions in matter above the
nuclear saturation density. In order to do this, we parametrize the
EoS of matter at higher densities with a simple expression rich enough
to include exotic matter. We perform a Bayesian analysis using data
from QLMXBs and neutron stars which exhibit PRE bursts, where our
model space is given by the EoS parameters and also one parameter for
the mass of each neutron star in the data set. Given an EoS, the TOV
equations provide the M-R curve and thus a prediction for the radius
of each neutron star from its mass. As described above, we always
ensure that our EoS are causal, hydrodynamically stable, and that our
M-R curves support a 2 $M_{\odot}$ neutron stars. 

It is important to note that these results are sensitive to several
model assumptions and also sensitive to the EoS parameterization that
we use. This is demonstrated in Table~\ref{tab:radii}, where the 68\%
and 95\% confidence limits are given for several different EoS
parametrizations (top portion of the table) and variations in the
interpretation of the data (bottom portion). The full specification of
the models and data modifications is given in
Ref.~\cite{Steiner:2013}. The first row (model A) is a baseline model
where the high-density part of the EoS was described with two
polytropes. The alternate EoS parameterization which most strongly
changes the radius is that of model C, which treats the EoS as a set
of line segments in the pressure-energy density plane. This allows for
very strong phase transitions, typical of that obtained in a Maxwell
construction where the pressure is very flat as a function of density.
Model D describes a hybrid neutron star with deconfined quark matter
at the core. In this case, the higher-density polytrope is replaced by
the quark matter model of Ref.~\cite{Alford:2005}.

\begin{table}
\begin{tabular}{llrrrr}
\hline
EoS model & Data modifications & 
$R_{95\%>}$ (km) & $R_{68\%>}$ (km) & 
$R_{68\%<}$ (km) & $R_{95\%<}$ (km) \\
\hline
\multicolumn{6}{c}{Variations in the EoS model} \\
A (2 polytropes) & & 11.18 & 11.49 & 12.07 & 12.33 \\
B (2 polytropes) & & 11.23 & 11.53 & 12.17 & 12.45 \\
C (line segments) & & 10.63 & 10.88 & 11.45 & 11.83 \\
D (w/quarks) & & 11.44 & 11.69 & 12.27 & 12.54 \\
\multicolumn{6}{c}{Variations in the data interpretation} \\
A & I (high $f_C$) & 11.82 & 12.07 & 12.62 & 12.89 \\
A & II (low $f_C$) & 10.42 & 10.58 & 11.09 & 11.61 \\
A & III ($z_{\mathrm{ph}}=z_{\mathrm{NS}}$) & 10.74 & 10.93 & 11.46 & 11.72 \\
A & IV (without X7) & 10.87 & 11.19 & 11.81 & 12.13 \\
A & V (without M13) & 10.94 & 11.25 & 11.88 & 12.22 \\
A & VI (no PREs) & 11.23 & 11.56 & 12.23 & 12.49 \\
A & VII (no qLMXBs) & 11.17 & 11.96 & 12.47 & 12.81 \\
Global limits & & 10.42 & 10.58 & 12.62 & 12.89\\
\hline
\end{tabular}
\caption{Limits for the radius of a 1.4 solar mass neutron star for
  several different EoS models and interpretations of the
  data.\label{tab:radii}}
\end{table}

The largest uncertainty in the radius is obtained from the variation
in $f_C$, the color correction factor. This factor describes the
deviation of the X-ray spectrum from a black-body during the ``cooling
tail'' of a PRE-burst. We also examine the variations in the radius
after having removed extreme neutron stars, or mass-radius
distributions obtained from QLMXBs or PRE X-ray bursts. Over all of
the changes we make to the EoS model and the interpretation of the
data, the radius of a 1.4$M_{\odot}$ neutron star lies between 10.4
and 12.9 km. Nevertheless, we have not tried all possible data
interpretations and EoS models. Such a task is impossible, simply
because there is no unambiguous way to enumerate a uncountably
infinite parameter space (similar to the result that the cardinality
of the real numbers is larger than that of the integers). Our choice
of EOS models and data interpretations is thus necessarily biased, and
this uncertainty is manifest as the prior distributions of our
Bayesian analysis.

\begin{figure}
\begin{center}
\includegraphics[width=0.6\textwidth]{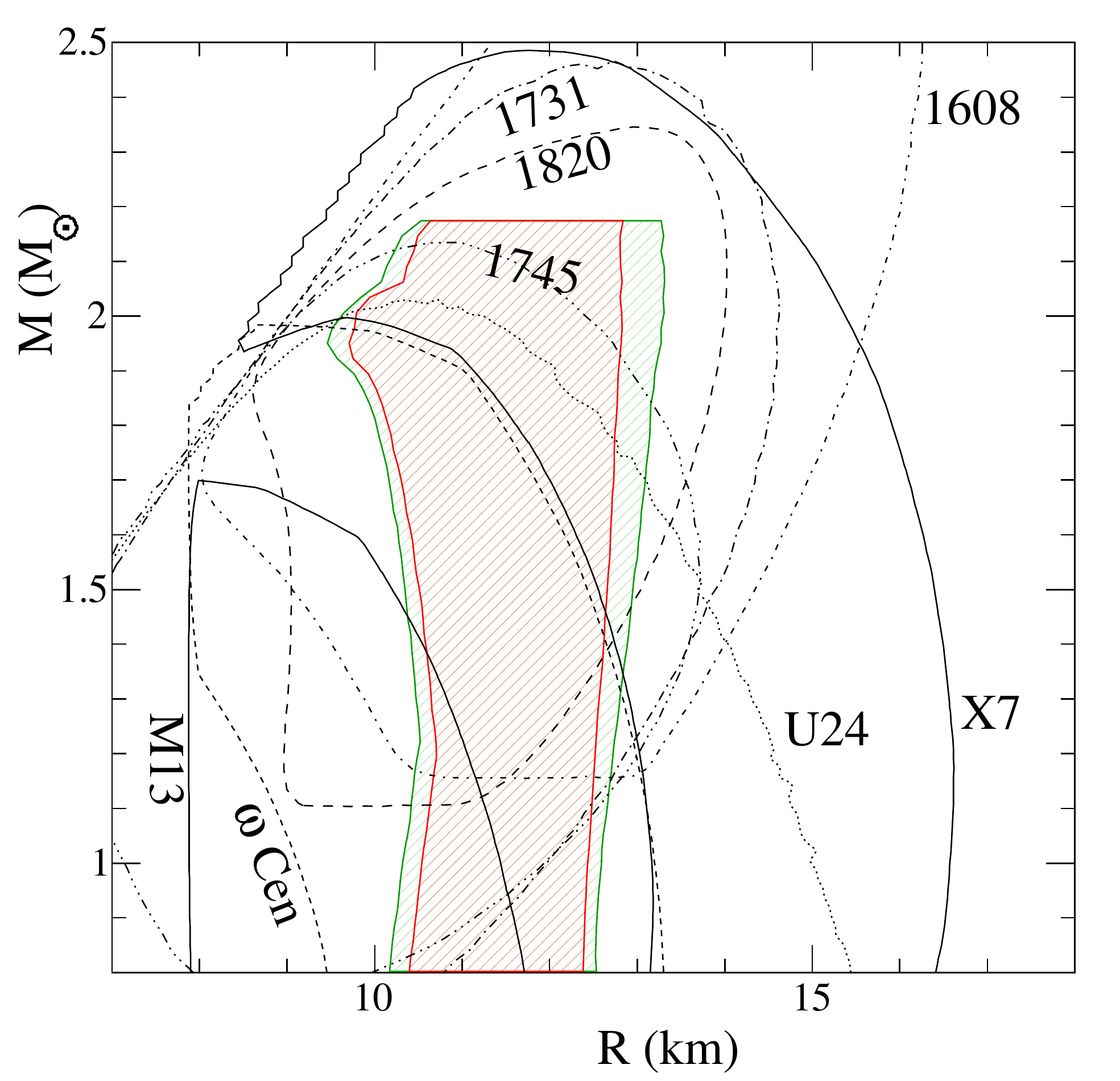}
\includegraphics[width=0.6\textwidth]{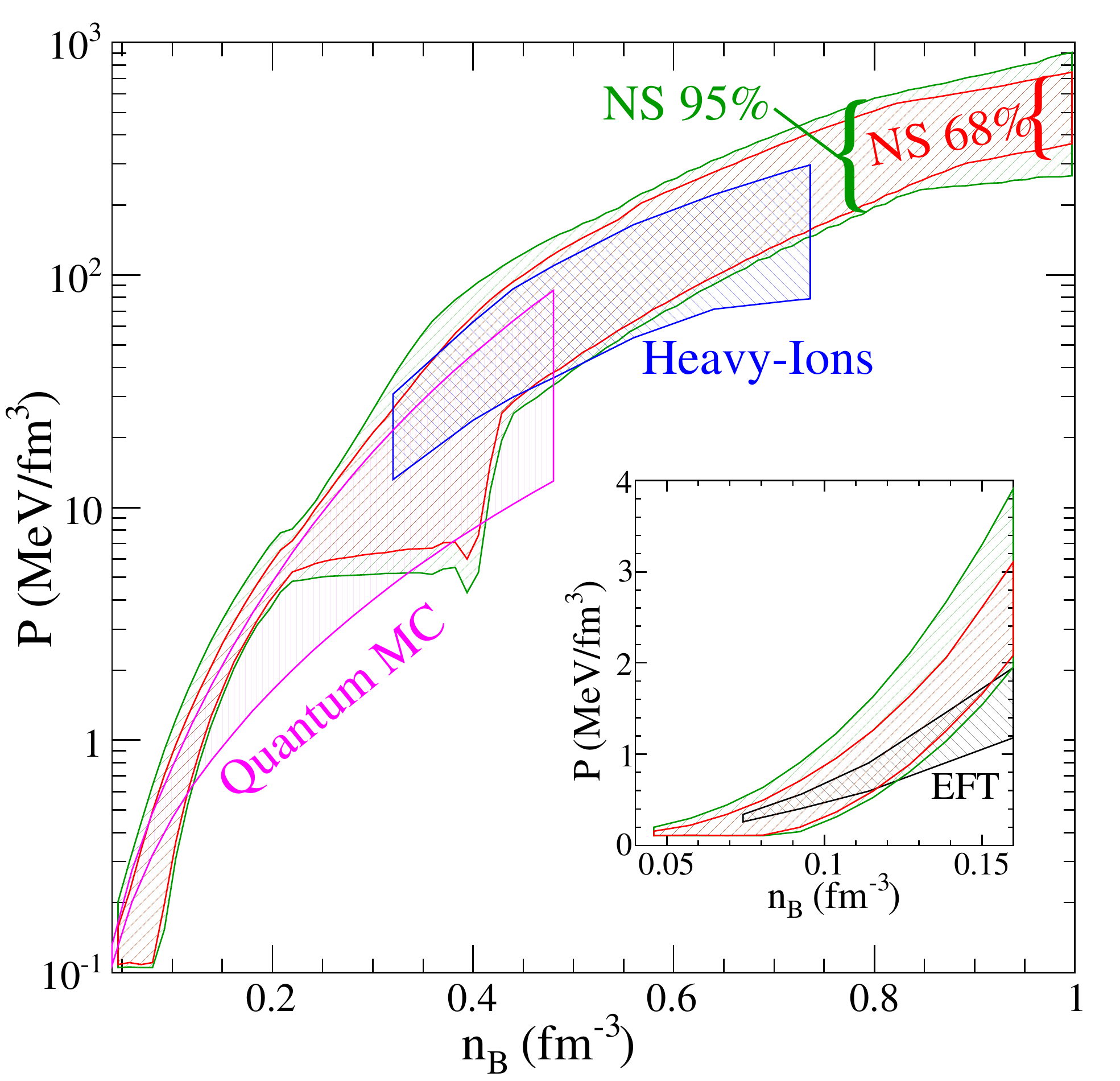}
\end{center}
\caption{ A comparison of the predicted $M\textrm{--}R$ relation with
  the observations. The shaded regions outline the 68\% and 95\%
  confidences for the $M\textrm{--}R$ relation; these include
  variations in the EoS model and the modifications to the data set
  (see Table~\ref{tab:radii}) but not the more extreme scenarios. The
  lines give the 95\% confidence regions for the eight neutron stars
  in our data set. The predicted pressure as a function of baryon
  density of neutron-star matter as obtained from astrophysical
  observations. The region labeled ``NS 68\%'' gives the 68\%
  confidence limits and the region labeled ``NS 95 \%'' gives the 95\%
  confidence limits. Results for neutron-star matter from effective
  field theory~\cite{Hebeler:2010} (see inset), from quantum Monte
  Carlo~\cite{Gandolfi:2012}, and from constraints inferred from
  heavy-ion collisions~\cite{Danielewicz:2002} are also shown for
  comparison.\label{fig:slb13} }
\end{figure}

The final results for the $M-R$ curve and EoS are given in
Fig.~\ref{fig:slb13} from Ref.~\cite{Steiner:2013}. The $M-R$ curve
obtained is relatively vertical, which naturally implies that almost
all neutron stars have approximately the same radius. The EoS obtained
from the mass and radius observations is also in concordance with
results from quantum Monte Carlo and chiral effective theory described
above and constraints obtained from heavy-ion collisions.

\section{Determining the Density Dependence of the Symmetry Energy}

In order to determine the symmetry energy, we use the parameterization
of the neutron matter EoS from the quantum Monte Carlo results in
Eq.~\ref{eq:enefunc} above. With this parameterization the symmetry
energy at the saturation density $E_{\mathrm{sym}}$ and the parameter
  which describes the density dependence of the symmetry energy, $L$,
  are given by
\begin{equation}
E_{\rm sym}=a+b+16 \,,\quad L=3\,(a\alpha+b\beta) \, .
\end{equation}
Neutron stars contain a small amount of
protons, so we multiply the EoS by a small ($\sim$ 10\%) and
density-dependent correction factor which modifies the pressure. This
correction factor is obtained by averaging over Skyrme forces which
give similar M-R curves to those suggested by the data. At some higher
density $\rho_t \sim 0.24 - 0.48$ fm$^{-3}$ the EoS may change due to
the presence of exotic matter or a higher-order many-body correction.
Beginning with this density, we employ a polytrope of the form $P=K_1
\epsilon^{\Gamma}$, fixing $K_1$ to ensure that the EoS is continuous
and setting $\Gamma_1=1+1/n_1$ where $n_1$ is the ``polytropic
index''. At a higher energy density, $\epsilon_2$, we use a second
polytrope with index $n_2$, fixing $K_2$ to ensure that the EoS is
continuous. This very similar to the ``model A'' described in
Sec.~\ref{sec:bayes} above. Generally, we find similar $M-R$ curves,
independent of whether or not the neutron star contains quark matter
in the core. We also find that the effect of varying $\rho_t$ between
0.24 and 0.48 fm$^{-3}$ is relatively small.

\begin{figure}
\begin{center}
\parbox{2.1in}{
\vspace*{-1.5in}
\includegraphics[width=2.1in]{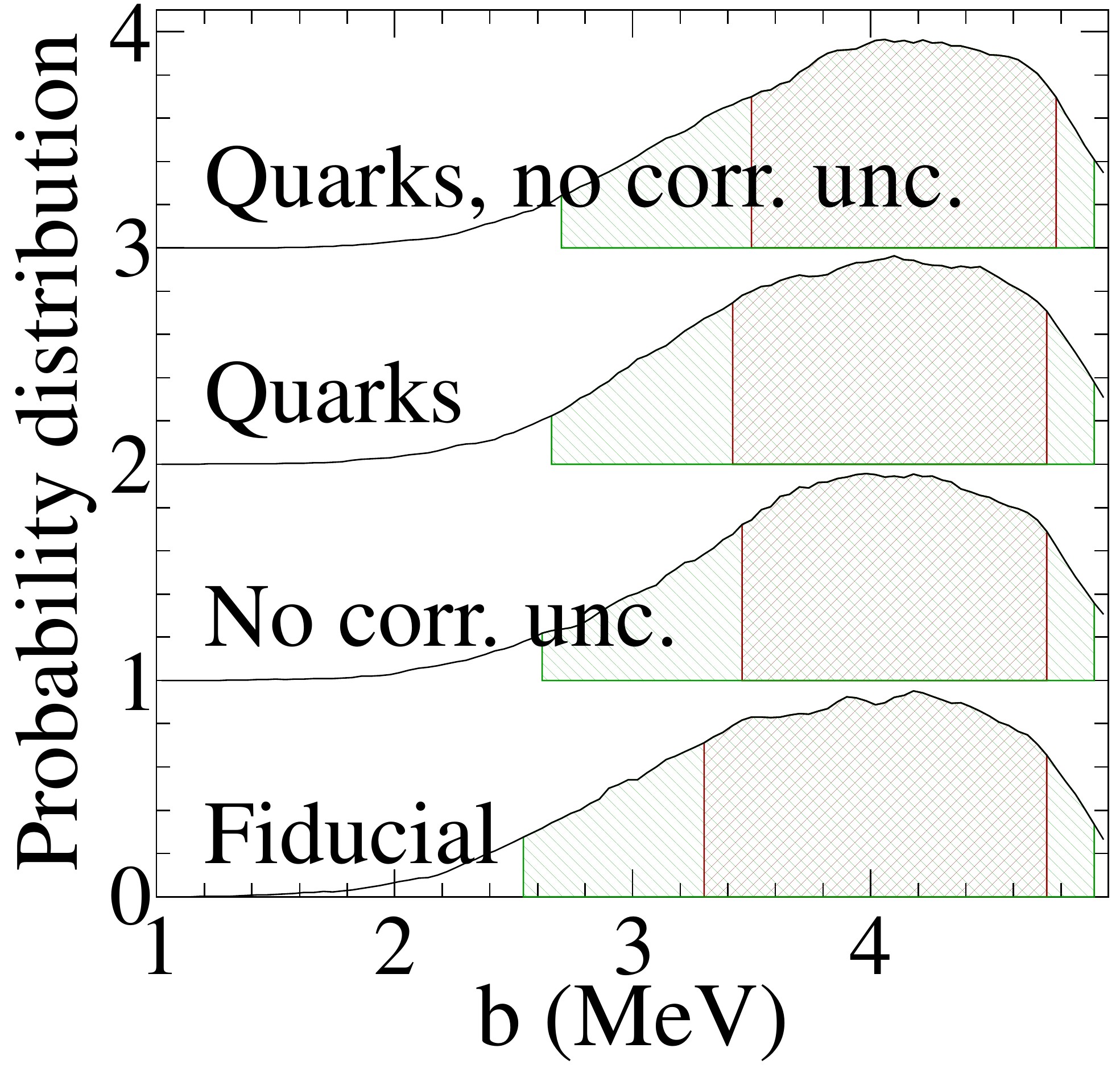}
\includegraphics[width=2.1in]{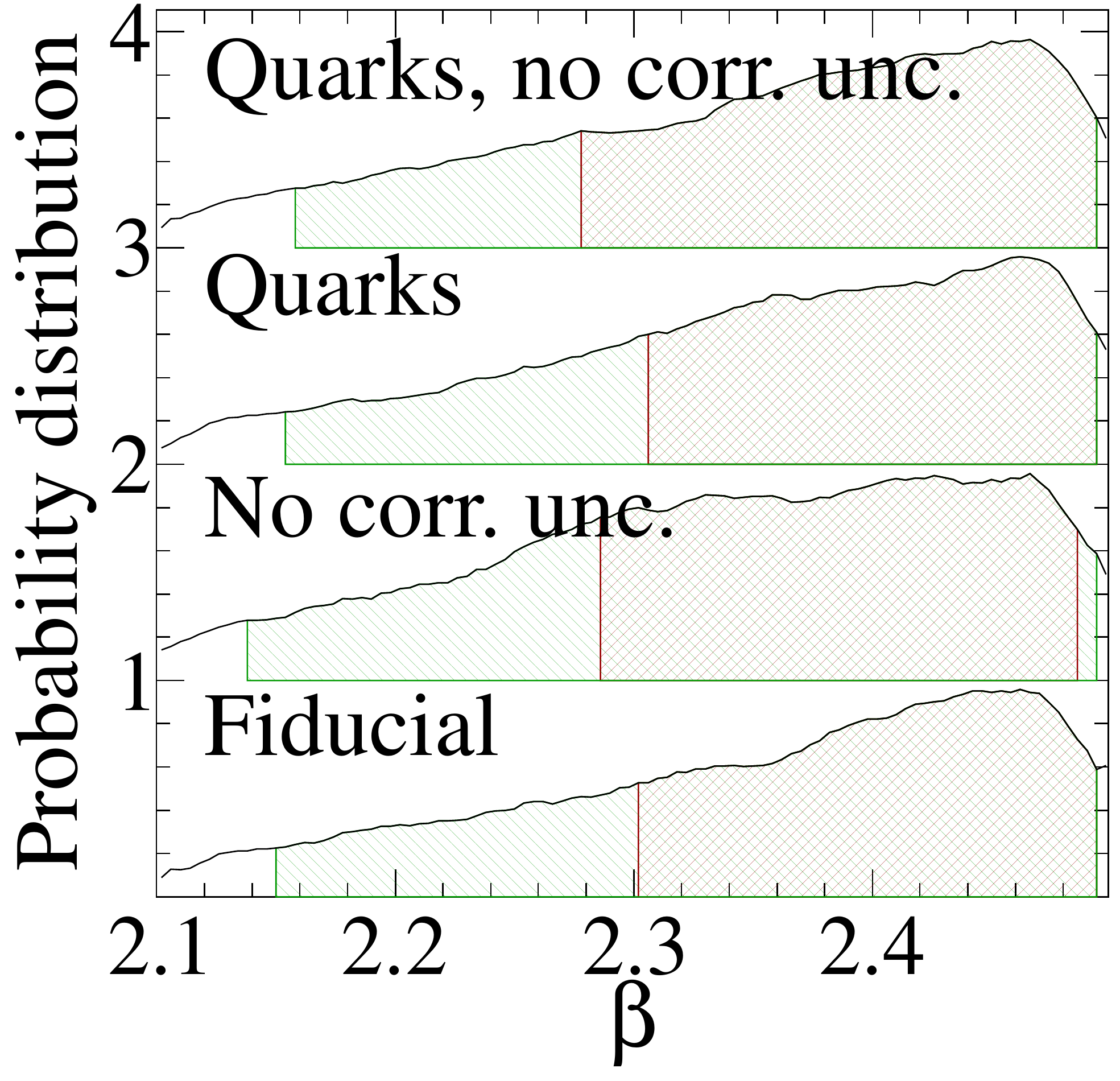}
}
\includegraphics[width=2.1in]{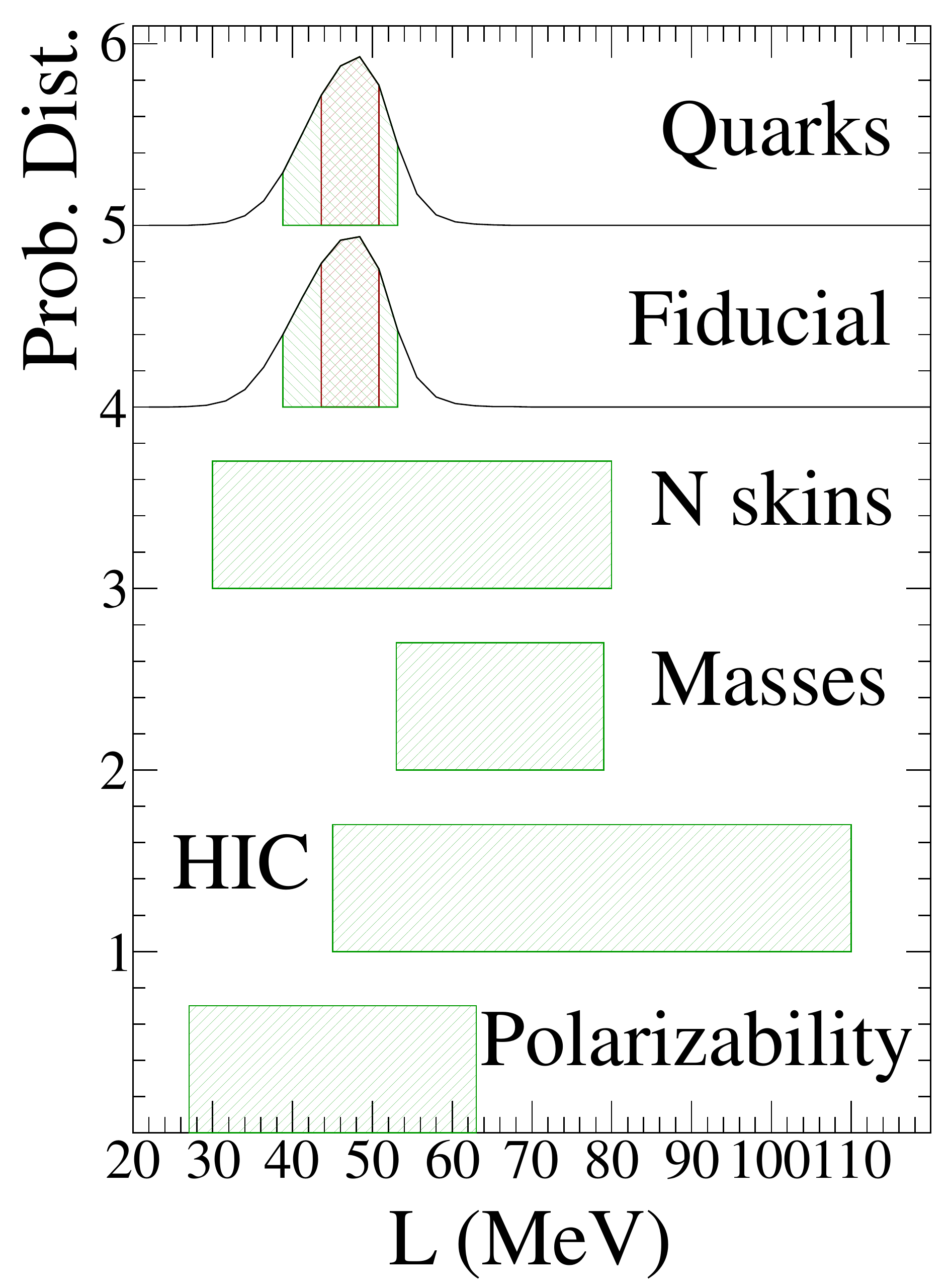}
\caption{The left panel shows probability distributions of the
  parameters $b$ and $\beta$ obtained from the Bayesian analysis. The
  right panel summarizes constraints on $L$ from observations and
  experiments. The top two curves show constraints on $L$ as
  probability distributions assuming either the fiducial model of
  Ref.~\cite{Steiner:2012} or the model containing quarks. The bottom
  four curves show constraints on $L$ from experiment, from neutron
  skins~\cite{Warda:2009}, nuclear masses~\cite{Liu:2010}, heavy-ion
  collisions~\cite{Tsang:2009}, and from the electric dipole
  polarizability~\cite{Tamii:2011}.}
\label{fig:params}
\end{center}
\end{figure}

This analysis also provides posterior probability distributions for
the EoS parameters. While we do not obtain significant constraints on
$a$ or $\alpha$, the mass and radius data do constrain the parameters
$b$ and $\beta$ (Fig.~\ref{fig:params}). While the simple
parametrization employed in this section cannot fully describe the
complexities of the nuclear three-body force, it does make it clear
that astrophysical data is beginning to rule out some three-body
forces which might otherwise be acceptable. We also show constraints
on $L$. From neutron stars we obtain the constraints to the symmetry
energy and slope to be $32<E_{\rm sym}<34~\mathrm{MeV}$ and $43<{\rm
  L}<52~\mathrm{MeV}$ within 68\% confidence. The only way to obtain a
larger value of $L$ is through a strong phase transition just above
the nuclear saturation density which tends to decouple the properties
of matter at low- and high-densities. Thus model C from
Sec.~\ref{sec:bayes} above allows values of $L$ as large as 83 MeV.
However, it is not clear that such a strong phase transition at low
densities is particularly realistic, as it might have been already
ruled out by experimental work in heavy-ion collisions as reviewed
in Ref.~\cite{Tsang:2012}.

\section{Acknowledgements}
SG is supported by the U.S.~Department of Energy, Office of Nuclear
Physics, by the NUCLEI SciDAC program and by the LANL LDRD program.
AWS is supported by DOE Grant No. DE-FG02-00ER41132.

\bibliography{biblio}

\end{document}